\documentstyle[amssymb,11pt]{article}

\input{tcilatex}
\input epsf
\begin{document}

\font\cmss=cmss10 \font\cmsss=cmss10 at 7pt 
\hfill CERN-TH/99-33

\hfill

\vspace{20pt}

\begin{center}
{\Large {\bf {Anomalies, Unitarity and Quantum Irreversibility}}}
\end{center}

\vspace{6pt}

\begin{center}
{\sl Damiano Anselmi}

{\it CERN, Division Th\'eorique, CH-1211, Geneva 23, Switzerland }
\end{center}

\vspace{12pt}

\begin{center}
{\bf Abstract}
\end{center}

\vspace{4pt} {\small The trace anomaly in external gravity is the sum of
three terms at criticality:\ the square of the Weyl tensor, the Euler
density and $\Box R$, with coefficients, properly normalized, called $c,$ $a$
and $a^{\prime }$, the latter being ambiguously defined by an additive
constant. Considerations
about unitarity and positivity properties of the induced actions allow
us to show that the total RG flows of $a$ and $a^{\prime }$ are equal and
therefore the $a^{\prime }$-ambiguity can be consistently removed through
the identification $a^{\prime }=a$. The picture that emerges clarifies
several long-standing issues. The interplay between unitarity and
renormalization implies that the flux of the renormalization group is
irreversible. A monotonically decreasing $a$-function interpolating between
the appropriate values is naturally provided by $a^{\prime }$. The total $a$%
-flow is expressed non-perturbatively as the invariant (i.e.
scheme-independent) area of the graph of the beta function between the fixed
points. We test this prediction to the fourth loop order in perturbation
theory, in QCD with $N_{f}\lesssim 11/2\,N_{c}$ and in supersymmetric QCD.
There is agreement also in the absence of an interacting fixed point (QED
and $\varphi ^{4}$-theory). Arguments for the positivity of $a$ are also
discussed.}

\vskip 7.5truecm \noindent CERN-TH/99-33 -- March, 1999

\vfill\eject 

\section{Introduction}

Anomalies are often calculable to high orders in perturbation theory with a
relatively moderate effort. Sometimes they are calculable exactly to all
orders.

Much of the present knowledge about the low-energy limit of asymptotically
free quantum field theories comes from anomalies, via the Adler--Bardeen
theorem \cite{bardeen}. Conserved axial currents have one-loop exact
anomalies and the 't Hooft anomaly matching conditions \cite{thooft} put
constraints on the low-energy limit of the theory.

A second class of anomalies, related to the stress tensor, called {\sl %
central charges}, do not satisfy the Adler--Bardeen theorem. Nevertheless,
they obey various positivity constraints, which also put restrictions on the
low-energy limit of the theory.

Other remarkable positivity constraints are those obeyed by the spectrum of
anomalous dimensions of the quantum conformal algebra, i.e. the algebra
generated by the operator product expansion of the stress tensor.
Applications of the Nachtmann theorem \cite{nach} reveal non-trivial
properties of strongly coupled conformal field theories \cite{n=4,n=2},
especially in the presence of supersymmetry, where the algebraic structure
simplifies considerably.

Furthermore, in supersymmetric theories, the two classes of anomalies
mentioned above, axial and trace, are related to each other, and the
Adler--Bardeen theorem can be used to compute the exact IR values of the
central charges in the conformal window \cite{noi}. The consequent large
class of restrictions on the low-energy limit of the theory can be studied
explicitly \cite{noi2}.

There are positivity properties of the central charges that have not been
rigorously proved, yet. The purpose of this paper is to clarify certain
long-standing issues in this context.

The trace anomaly of the energy-momentum tensor is deeply related to the
renormalization group flow. There is empirical evidence \cite{noi,noi2} that
a central charge, called $a$, is positive and takes greater values in the UV
than in the IR: $a_{{\rm UV}}\geq a_{{\rm IR}}\geq 0$. The quantity $a$ is
interpreted as the number of massless degrees of freedom of the theory. This
means that the flux of the renormalization group is irreversible.

We call this notion {\sl quantum} irreversibility, to distinguish it from
time irreversibility, proper of thermodynamics and statistical mechanics.

A first suggestion in favour of this idea comes from two-dimensional quantum
field theory, where Zamolodchikov \cite{zamolo} proved that the central
extension $c$ of the stress tensor operator product algebra is positive and
monotonically decreasing along the renormalization group flow.

A four-dimensional generalization of this property is, however, more
difficult to prove. In four dimensions, for example, the set of candidate
central charges is richer, and among them there is also the central
extension $c$ of the operator product algebra. Various proposals for the
good candidate have appeared in the literature, as well as attempts to prove
the irreversibility property. We do not review the history of this research
here, but one proposal, due to Cardy \cite{cardy}, deserves special mention,
since the results of \cite{noi} were able to reject all the other
candidates, in particular the central extension $c$ of the OPE algebra. At
the same time, the impressive amount of evidence in favour of the ``$a$%
-theorem'' \cite{noi,noi2} convinced many people that quantum
irreversibility was true.

In this paper we reconsider the matter under a different viewpoint. We
present a general picture that clarifies various issues related with the
trace anomaly and unifies some notions that have been so far considered
unrelated. Quantum irreversibility is intrinsically contained in this
picture, and the outcome is an explicit non-perturbative formula for the
total flow of $a$ that can be tested successfully to the fourth loop order
in perturbation theory. This formula gives an intuitive (and geometrical)
picture of quantum irreversibility, measured by the area of the graph of the
beta function.

In the rest of the introduction we give the basic guidelines of our
arguments, anticipate some applications, and explain how the paper is
structured. The presentation is meant to be self-consistent and we take this
opportunity to discuss the same issue under different viewpoints.

In four dimensions the trace anomaly operator equation in an external
gravitational field can be written in the form 
\begin{equation}
\Theta =-\frac{1}{120}\frac{1}{(4\pi )^{2}}\left[ \tilde{c}(\alpha )\ W^{2}-%
\frac{1}{3}\tilde{a}(\alpha )\ {\rm G}+\frac{2}{9}\tilde{a}^{\prime }(\alpha
)\ \Box R+\beta (\alpha )h(\alpha )\ R^{2}\right] +\frac{1}{4}\ \beta
(\alpha )F^{2},  \label{t}
\end{equation}
where $\alpha $ denotes the renormalized coupling constant, at a certain
reference scale $\mu $, and we have defined the $\beta $-function as 
\begin{equation}
\beta (\alpha )=\frac{1}{\alpha }\mu \frac{{\rm d}\alpha }{{\rm d}\mu }=%
\frac{{\rm d\ln }\alpha }{{\rm d\ln }\mu };  \label{see}
\end{equation}
$W$ is the Weyl tensor of the external gravitational field, $W^{2}=R_{\mu
\nu \rho \sigma }R^{\mu \nu \rho \sigma }-2R_{\mu \nu }R^{\mu \nu }+\frac{1}{%
3}R^{2}$, and $G=R_{\mu \nu \rho \sigma }R^{\mu \nu \rho \sigma }-4R_{\mu
\nu }R^{\mu \nu }+R^{2}$ is the Gauss--Bonnet integrand. The stress tensor
is $T_{{}}^{\mu \nu }=-\frac{2}{\sqrt{-g}}\ \frac{\delta S}{\delta g_{\mu
\nu }^{{}}},$ $S$ denoting the action in the gravitational background. The
curvature conventions are those of refs. \cite{hathrell,hathrell2}. We work
partly in the Lorentzian framework and partly in the Euclidean framework.

The last term of (\ref{t}) is written, for concreteness, in the case of
Yang--Mills theory. In general it should read $\sum_{i}\beta _{i}{\cal O}%
_{i} $, where the sum runs over the set of coupling constants of the theory.
In the presence of scalar fields $\varphi $ there is an additional
complication, which has to be treated apart, due to the renormalization
mixing between the stress-energy tensor and the operator $\left( \partial
_{\mu }\partial _{\nu }-\eta _{\mu \nu }\Box \right) \varphi ^{2}$, $\eta
_{\mu \nu }={\rm diag}(1,-1,-1,-1)$. Arguments and conclusions are valid for
the most general renormalizable quantum field theory.

The term $R^{2}$ does not contribute at criticality, since it is neither a
total derivative nor a conformal-invariant. For this reason one can
factorize a $\beta (\alpha )$ in its coefficient, which multiplies a certain
function $h(\alpha )$.

The tilde over the functions $\tilde{c}(\alpha )$, $\tilde{a}(\alpha )$ and $%
\tilde{a}^{\prime }(\alpha )$ is used to remark that these functions, as
they appear in the anomaly operator equation, have not a direct physical
meaning at a generic energy scale (see the discussion in section 1 of ref. 
\cite{ccfis} for all details). In particular, they are scheme-dependent.
Physical quantities have to be defined via matrix elements of operators
rather than the operator equations. An operator equation contains artefacts
that disappear in matrix elements. In practice, when inserting eq. (\ref{t})
inside correlators, the contributions coming from the matrix elements of the
dynamical operator ${\frac{\beta }{4}}F^{2}$ will restore full scheme
independence. Only at criticality, where $\beta =0$, is there no such
ambiguity, where $\tilde{c}_{*}$ and $\tilde{a}_{*}$ have a direct physical
meaning. Instead, $a_{*}^{\prime }$ retains a peculiar type of ambiguity
that we are going to discuss in detail.

In order to properly interpolate between the UV\ and IR\ critical values,
one has to define physical (i.e. scheme-independent) {\it central functions}%
, $c(\alpha )$, $a(\alpha )$ and $a^{\prime }(\alpha ),$ through matrix
elements of operators. In ref. \cite{ccfis} this was done for the function $%
c(\alpha )$ and certain ``secondary'' central charges. In the first part of
the paper we extend this analysis to the function $a^{\prime }(\alpha )$ and
prove that it satisfies the irreversibility property. In the rest of the
paper we explain how this interpolating function is also a good central
function for $a(\alpha )$, so that $a$ satisfies the irreversibility
property as well (the ``$a$-theorem'', see \cite{noi}).

Reflection positivity implies 
\[
c\geq 0, 
\]
since $c$ is the overall constant of the stress tensor two-point function,
whose structure is uniquely fixed by conformality at the fixed points. At
the OPE level \cite{n=4,n=2} $c$ represents the central extension of the
quantum conformal algebra, which is the reason why we retain the symbol $c$
for it. It has been shown, even at the non-perturbative level \cite{noi} in
the conformal window, that the central extension is not, in general, the
quantity that monotonically decreases along the RG\ flow: this is true only
in two dimensions. Following \cite{noi} we use a different symbol for the
decreasing quantity, $a$, and speak of ``$a$-theorem''. At the level of OPE
algebra, the interpretation of the quantity $a$ is different from that in
two dimensions \cite{n=2}: the combination $1-a/c$ is indeed a structure
constant of the OPE algebra. 

There are cases, also in four dimensions, where the central extension does
decrease from the UV\ to the IR, for example when the theory interpolates
between conformal fixed points with $c=a$. These conformal field theories
have a simplified OPE\ algebra and other nice properties \cite{n=4,n=2}. $c$
decreases in several other particular models, or in part of the conformal
window. Examples of this kind are treated in \cite{noi,noi2}. Nevertheless,
this behaviour is not general and so the central extension is not a good
counter of the massless degrees of freedom of the system.

The quantity $a$ has been shown to have the desirable properties, namely 
\begin{equation}
a_{{\rm UV}}\geq a_{{\rm IR}}\geq 0,
\end{equation}
in various concrete models, the most impressive results being the exact
formulas of \cite{noi} for the conformal windows in supersymmetric theories,
applied in \cite{noi2} to a large variety of cases. Off the conformal
window, non-perturbative tests based on general physical grounds in QCD \cite
{cardy} and on various duality conjectures in supersymmetric theories \cite
{bastianelly} are successful, but less constraining, since $c$ passes them
also. At the rigorous level, both proofs that $a$ is positive and decreasing
along the RG\ flow have been missing. Positivity of $a$ passes the tests of 
\cite{noi,noi2} within the known conformal windows. The breakdown of this
condition typically signals that the IR\ fixed point does not exist (as in
pure N=1 supersymmetric QCD).

Finally, the quantity $a^{\prime }$ has remained, up to now, somewhat
mysterious, especially at criticality; yet it is simple to prove that it is
monotonically decreasing from the UV\ to the IR. There exists a central
function $a^{\prime }(t)=a^{\prime }[\alpha (t)]$ that satisfies the
irreversibility property at any intermediate energy scale, 
\[
{\rm d}a^{\prime }(t)/{\rm d}t\leq 0; 
\]
nevertheless, as it stands, $a^{\prime }$ is not a good counter of degrees
of freedom, since it is meaningless at criticality. One of the purposes of
our analysis is to clarify the meaning of $a^{\prime }$ and its relationship
with the other two quantities, in particular the quantity $a$.

We see that each of the three quantities has part of the properties that we
would like for a single quantity: $c$ is positive, but not monotonically
decreasing; $a^{\prime }$ is monotonically decreasing, but ill-defined at
criticality; $a$ has the good properties, but so far only at the empirical
level, in the sense that they have not been proved rigorously. It is only by
uncovering the deep meaning of each quantity and the interplay among them
that a clarifying picture can emerge. We can say that the matter is much
simpler in two dimensions, because, in some sense, ``$c=a=a^{\prime }$''.
The proof of irreversibility \cite{zamolo} and positivity are
straightforward in two dimensions.

In some works \cite{birreldavies} one finds arguments in favour of the
identification $a^{\prime }=3c$. This is however an artefact of the
regularization scheme (a dimensional continuation preserving conformal
invariance in $d$-dimensions) and is actually inconsistent. Indeed, if this
equality were consistent, it would hold both in the UV limit and the IR
limit. However, $a^{\prime }$ is monotonically decreasing, while $c$ has an
indefinite behaviour, as proved in \cite{noi}. There are many known examples
where $a^{\prime}_{{\rm UV}}=3~ c_{{\rm UV}}$ does not imply $a^{\prime}_{%
{\rm IR}}=3~ c_{{\rm IR}}$.

Therefore, if the ambiguous quantity $a^{\prime}$ has to be identified with
one of the two unambiguous central charges, or a linear combination of them,
it can only be identified with $a$. The relative factor can be chosen in
such a way that the relation $a^{\prime}=a$ has other noticeable properties.
In particular the induced action for the conformal factor (the Riegert
action \cite{riegert}) simplifies enormously (it becomes free as in two
dimensions).

Considerations about positivity of induced effective actions (absence of
negatively normed states) allow us to show that the identification $%
a^{\prime}=a$ is consistent, i.e. that if assumed in the UV limit of the
theory it holds also in the IR limit, precisely 
\begin{equation}
a_{{\rm UV}}-a_{{\rm IR}}=a_{{\rm UV}}^{\prime }-a_{{\rm IR}}^{\prime }.
\label{rr}
\end{equation}
The ambiguity of the quantity $a^{\prime }$ can be resolved by fixing $a_{%
{\rm UV}}^{\prime }=a_{{\rm UV}}.$ Then (\ref{rr}) implies $a_{{\rm IR}%
}^{\prime }=a_{{\rm IR}}.$ Therefore, according to the above observations,
one can define a monotonically decreasing physical function $a^{\prime
}[\alpha (t)]$ at all intermediate energies, whose values coincide with the
values of $a$ at the critical points. In this interpretation
irreversibility is the result of the
interplay between unitarity and renormalization.

The same considerations show that $a$ is positive throughtout the RG\ flow,
once it is positive at some reference energy and since $a>0$ in the free
field limit, we have $a\geq 0$ also in the interacting fixed point.

From the identification $a^{\prime}=a$, a simple non-perturbative formula
expressing the total RG flow of $a$ as the invariant area of the graph of
the beta function follows (here ``invariant'' area means
scheme-independent). This formula can be checked in perturbation theory, to
the fourth loop order included, in all renormalizable models.

Using our results, a notion of ``proper'' coupling constant $\bar{\alpha}$
can be defined, which is the coupling constant for which the
``Zamolodchikov'' metric is constant throughtout the renormalization group
flow. The total flows of $a$ and $a^{\prime }$ equal the area of the graph
of the proper beta function (i.e. the beta function for $\bar{\alpha}$)
between the fixed points. This area is quantized in QCD. In general, one can
say that a universal unit-area cell is assigned to each massless degree of
freedom.

The paper is organized into two main parts. The first part (section 2) is
devoted to the interpolating function for $a^{\prime }$, the second part
(section 3) to the removal of its ambiguity, through the relationship with $%
a $.

Other implications are presented in the final part of the paper and the
conclusions. In particular, we discuss the induced action for the conformal
factor (the Riegert action) and show that our identification $a^{\prime }=a$
reduces it to a free action at criticality. Moreover, we derive an
expression for the vacuum energy $E_0$.

A comment on the claimed irreversibility is in order. The statement is about
the {\it intrinsic} irreversible character of the flux of the
renormalization group. By intrinsic we mean proper to the dynamical scale $%
\mu $ introduced by renormalization. The desired effect has to be suitably
``cleaned'' from spurious effects of different nature, that can either
enhance or spoil the property in a trivial way. For this reason we consider
the most general renormalizable theory with no mass parameter. A\ mass would
trivially enhance the theorem, by killing degrees of freedom in the IR,
without modifying the UV. Instead, a non-renormalizable interaction would
trivially spoil the theorem, by killing degrees of freedom in the UV,
without modifying the IR. Even the non-perturbative effects of QCD, such as
chiral symmetry breaking, enhance the theorem, so that the crucial region
for testing the intrinsic irreversibility of the RG\ flow is precisely the
conformal window.




















\section{The quantity $a^{\prime }$: ambiguity, irreversibility,\newline
interpolating function}

The quantity $a^{\prime }$ is known to be ambiguous by an arbitrary additive
constant. A regularization technique can often hide this ambiguity and give
an apparently unambiguous result for $a^{\prime }$. For example, in \cite
{birreldavies}, $a^{\prime }$ is related to $c$. A general calculation can
be found in \cite{hathrell}, with relevant comments in the concluding
paragraph. This ambiguity, related to the addition of an arbitrary finite $%
\int R^{2}$-term in the induced effective action, does not spoil the $%
a^{\prime}$-physical content completely. For example, the $a^{\prime}$-RG
flow is unambiguously defined; $a^{\prime }$ is like an additional coupling
constant of the theory. Once it is normalized at a reference energy scale,
it is fixed at any other energy scale. Positive definiteness of the induced
effective action for the conformal factor imposes nevertheless a bound on $%
a^{\prime }$ (see section 3.2).

At criticality formula (\ref{t}) becomes 
\begin{equation}
\Theta =-\frac{1}{120}\frac{1}{(4\pi )^{2}}\left[ c_{*}\ W^{2}-\frac{1}{3}%
a_{*}\ {\rm G}+\frac{2}{9}a_{*}^{\prime }\ \Box R\right] .  \label{critica}
\end{equation}
For a free theory with $N_{s}$ real scalar fields, $N_{f}$ Dirac fermions,
and $N_{v}$ vector fields, we have \cite{birreldavies} 
\begin{equation}
c_{{\rm free}}=N_{s}+6N_{f}+12N_{v},\quad \quad a_{{\rm free}%
}=N_{s}+11N_{f}+62N_{v},  \label{free}
\end{equation}
while $a_{{\rm free}}^{\prime }$ remains for the moment undetermined. For
the central charges we use over-all normalizations different from those of
ref. \cite{noi} and the previous literature, in order to have integer valued
quantities at the free critical points.

The operator $\Theta $ is associated with the conformal factor $\phi $ of
the metric, $g_{\mu \nu }={\rm e}^{2\phi }\delta _{\mu \nu }$. $W$ does not
depend on $\phi $, while the Euler density depends on $\phi $ quadratically.
Instead, the term $\Box R$ contains a linear term in the conformal factor $%
\phi $ around flat space, $R=-6\,{\rm e}^{-2\phi }\left[ \Box \phi
+(\partial _{\mu }\phi )^{2}\right] $. Therefore, the two-point function of $%
\Theta $ is proportional to the number $a^{\prime}_{*}$ in the conformal
limit $\beta =0$. Using $\Theta =-{\rm e}^{-4\phi }\delta S/\delta \phi $ we
have 
\begin{equation}
\langle\Theta (x)\ \Theta (y)\rangle=i \left.\frac{\delta \langle\Theta
(x)\rangle}{\delta \phi (y)}\right|_{\phi=0}=-i\left.\frac{\delta^2 S_{{\rm %
eff}}[\phi]}{\delta \phi (x)\delta \phi (y)} \right|_{\phi=0}= \frac{i}{%
90(4\pi )^{2}}a_{*}^{\prime }\ \Box ^{2}\delta (x-y).  \label{aprimo}
\end{equation}
Here $S_{{\rm eff}}[\phi]$ denotes the induced effective action for the
conformal factor (it will be calculated explicitly in sect. 3.1). Turning to
the Euclidean framework, we get 
\[
\left.\langle\Theta (x)\ \Theta (y)\rangle\right|_{{\rm E}}= - \left.\frac{%
\delta^2 S_{{\rm E}}[\phi]}{\delta \phi (x)\delta \phi (y)}
\right|_{\phi=0}= -\frac{1}{90(4\pi )^{2}}a_{*}^{\prime }\ \Box ^{2}\delta
(x-y). 
\]
The subscript E denotes correlators and quantities in the Euclidean frame. A
positive definite effective action implies $a^{\prime }>0$ and the negative
sign in (\ref{aprimo}) is consistent with this. In two dimensions $%
\langle\Theta(x)\,\Theta(0)\rangle=-c{\frac{\pi}{3}}\Box\delta(x)$.

Using formula (\ref{aprimo}) we see that the quantity $a_{*}^{\prime }$ can
be expressed at criticality by the integral 
\begin{equation}
a_{*}^{\prime }=-\frac{15}{2}\pi ^{2}\int {\rm d}^{4}x\
|x-y|^{4}\langle\Theta (x)\ \Theta (y)\rangle_{{\rm E}}.  \label{ugh}
\end{equation}

\subsection{Generalities about the $a^{\prime }$-function}

We now consider the off-critical theory. We can define a function $a^{\prime
}(r)$ of the intermediate energy scale $1/r$ by restricting the integration
over a four-sphere $S(r,y)$ of radius $r$ and centred at the point $y$,
precisely 
\begin{equation}
a^{\prime }(r_{2})-a^{\prime }(r_{1})=-\frac{15}{2}\pi
^{2}\int_{S(r_{1},y)}^{S(r_{2},y)}{\rm d}^{4}x\ |x-y|^{4}\langle\Theta (x)\
\Theta (y)\rangle.  \label{ar}
\end{equation}
The notation means that the integral is performed in the region contained
between the two four-spheres. Unless differently specified, correlators are
in the Euclidean framework. Our formula (\ref{ar}) does not give the
critical values 
\[
a_{{\rm UV}}^{\prime }=\lim_{r\rightarrow 0}a^{\prime }(r),\quad \quad a_{%
{\rm IR}}^{\prime }=\lim_{r\rightarrow \infty }a^{\prime }(r) 
\]
of the function $a^{\prime }$, which are related to the ambiguity already
mentioned. Nevertheless, we prove in this paper that there is also a
universal way to remove it.

For a critical theory we have, from formula (\ref{aprimo}), $a_{{\rm UV}%
}^{\prime }=a_{{\rm IR}}^{\prime }=a^{\prime }(r)=a_{*}^{\prime }.$
Off-criticality, the running of $a^{\prime }(r)$ is due to the internal term 
$\frac{\beta }{4\alpha }F^{2}$ appearing in the operator anomaly equation (%
\ref{t}) for $\Theta $. Its effect is a non-local term in the correlator $%
\langle\Theta (x)~\Theta (y)\rangle$, which we write as 
\begin{equation}
\langle \Theta (x)\ \Theta (y)\rangle ={\frac{1}{15\pi ^{4}}}\,\,\frac{\beta
^{2}[\alpha (t)]f[\alpha (t)]}{|x-y|^{8}}~,~~~~~~~~~~{\rm for}~~x\neq y.
\label{tetateta}
\end{equation}
The function $f(t)$ has finite UV\ and IR\ limits. The regularity of the
function $f(t)$ at criticality follows from the very definition of the $%
\beta $-function and the operator $F^{2}$: the coefficients of the operators 
${\cal O}_{i}$ (i.e. the $\beta $-functions) are precisely the zeros of $%
\Theta $ in the operator equation $\Theta =\beta _{i}{\cal O}_{i}$ (for
example $\Theta =\frac{\beta }{4}F^{2}$). The proof can be found in the
classical works by Adler, Collins and Duncan \cite{adler}, Nielsen \cite
{nielsen} and Collins, Duncan and Joglekar \cite{joge} \footnotemark
\footnotetext{The existence of trace anomalies was first established
by Coleman and Jackiw in ref. \cite{CJ}.}. From this very fact
it follows, among the other things, that at criticality the anomalous
dimension of the operator $F^{2}$ coincides with the slope $\beta
_{*}^{\prime }$ of the $\beta $-function \cite{noiprimo}. For our purposes,
we just need that the function $f[\alpha (t)]$ be bounded and non-vanishing
at both critical points.

Reflection positivity \cite{osterwalder} of the correlator (\ref{tetateta})
at $x\neq y$ points assures that 
\begin{equation}
f(t)\geq 0.  \label{positivity}
\end{equation}

Now, let us insert (\ref{tetateta}) into (\ref{ar}). We obtain 
\begin{equation}
a^{\prime }(r)-a_{{\rm UV}}^{\prime }=-\frac{1}{2\pi ^{2}}\int_{S(r,y)}{\rm d%
}^{4}x\ \frac{\beta ^{2}(t)f(t)}{|x-y|^{4}}=-\int_{\alpha _{{\rm UV}%
}}^{\alpha (r)}\frac{{\rm d}\alpha }{\alpha }\ \beta (\alpha )f(\alpha
)=-\int_{-\infty }^{\ln r\mu }{\rm d}t\ \beta ^{2}(t)f(t),  \label{dir}
\end{equation}
and, for the total flow of the quantity $a^{\prime },$%
\begin{equation}
a_{{\rm UV}}^{\prime }-a_{{\rm IR}}^{\prime }=\int_{-\infty }^{+\infty }{\rm %
d}t\ \beta ^{2}(t)f(t)=-\int_{\alpha _{{\rm UV}}}^{\alpha _{{\rm IR}}}\frac{%
{\rm d}\alpha }{\alpha }\ \beta (\alpha )f(\alpha )\geq 0.  \label{resu}
\end{equation}
The integral is convergent\footnotemark 
\footnotetext{
Integrals like (\ref{ugh}) are the matter-induced gravitational couplings.
Specifically, $\int {\rm d}^4x\, \langle \Theta(x)\,\Theta(0)\rangle $, $%
\int {\rm d}^4x\, |x|^2 \langle \Theta(x)\,\Theta(0)\rangle $ and $\int {\rm %
d}^4x\, |x|^4 \langle \Theta(x)\,\Theta(0)\rangle $ are the induced
cosmological constant, the induced Newton constant and an induced
higher-derivative coupling, respectively \cite{adler2}. As they stand, the
first two integrals, however, are divergent \cite{cardy,zee}. No statement
can be made about the signs of the induced cosmological constant and the
Newton constant \cite{adler2,zee}. In particular, arguments for
irreversibility based on $\int {\rm d}^4x\, \langle
\Theta(x)\,\Theta(0)\rangle $ \cite{cardy,FL} present several unresolved
problems \cite{cardy}.} \cite{zee}. Around the UV\ fixed point (which we
assume to be free for concreteness) we have $\beta \sim -\alpha $, $\alpha
\sim -1/t$, $f\sim $ const., and $a^{\prime}_{{\rm UV}}-a^{\prime}_{{\rm IR}%
}\sim \int {\rm d}t/t^{2}$ convergent for large $t$. Around the IR\ fixed
point we have $\beta \sim \beta _{*}^{\prime }(\alpha -\alpha _{{\rm IR}}),$ 
$\beta _{*}^{\prime }>0.$ Solving the renormalization group equation, we
find $\beta \sim e^{-t\alpha _{{\rm IR}}\beta _{*}^{\prime }}$ \cite
{noiprimo}. As expected, convergence is much faster around the IR\ fixed
point, where it is exponential. In ref. \cite{zee}, Zee shows that
convergence holds also when $\beta\sim (\alpha-\alpha_*)^n$, $n>1$.

The monotonically decreasing behaviour of the function $a^{\prime }(r)$ is
evident: 
\[
{\frac{{\rm d}a^{\prime}(t)}{{\rm d} t}}=-\beta^2(t)f(t)\leq 0. 
\]

Summarizing, the integral expressing the total flow is convergent, if there
exists an IR fixed point $\alpha _{{\rm IR}}<\infty $ and it can be
interpreted as the {\it area} of the graph spanned by the beta function
between the fixed points. Since the $\beta $-function depends on the
subtraction scheme, while the area in question must be scheme independent,
the integral has to be performed with a suitable metric, an ``ein-bein''
that restores scheme invariance. This metric is precisely the function $%
f(\alpha )$.

\subsubsection*{An alternative expression}

For later convenience it is useful to re-express the correlator (\ref
{tetateta}) in a slightly different way, namely 
\begin{equation}
\langle\Theta (x)\ \Theta (y)\rangle=\frac{4}{45}\frac{1}{(4\pi )^{4}}\Box
^{2}\left( \frac{\beta ^{2}[\alpha (t)]\tilde{f}[\alpha (t)]}{|x-y|^{4}}%
\right) ~~~~~~~~~~~~~{\rm for}~~x\neq y.  \label{tetateta'}
\end{equation}
The above factorization of $\Box ^{2}$ comes naturally, for example, if one
writes the stress tensor two-point function as in formula (1.1) of ref. \cite
{ccfis}. One has 
\begin{equation}
\beta ^{2}(t)f(t)=\frac{1}{192}\left( {\frac{{\rm d}}{{\rm d}t}}-2\right)
\left( {\frac{{\rm d}}{{\rm d}t}}-4\right) ^{2}\left( {\frac{{\rm d}}{{\rm d}%
t}}-6\right) \beta ^{2}(t)\tilde{f}(t).  \label{dd}
\end{equation}

As far as positivity is concerned, we can safely cross the $\Box ^{2}$ and
infer positivity of $\tilde{f}$ by the positivity of $f$. This can be proved
via the following general argument. Let $v[\alpha (t)]$ and $u[\alpha (t)]$
be functions related by the equation 
\[
\left( \frac{{\rm d}}{{\rm d}t}-n\right) v[\alpha (t)]=u[\alpha (t)], 
\]
$n$ being a positive integer. Then we can prove that if $u$ is positive, $v$
is negative, and vice versa. Indeed, the solution of the differential
equation is 
\[
v\,[\alpha (t)]=-\int_{t}^{\infty }{\rm e}^{n(t-t^{\prime })}\,\,u[\alpha
(t^{\prime })]\,\,{\rm d}t^{\prime }. 
\]
The arbitrary constant is fixed by the requirement that $nv+u=0$ (i.e. ${\rm %
d}/{\rm d}t\equiv 0$) at criticality. This equality can be verified for,
say, $t\rightarrow -\infty $ by writing 
\begin{eqnarray*}
v\,[\alpha (t)] &=&\int_{0}^{-\infty }{\rm e}^{n\xi }\,\,u[\alpha (t-\xi
)]\,\,{\rm d}\xi \\
&&_{\overrightarrow{t\rightarrow -\infty }}\quad u[\alpha (-\infty
)]\int_{0}^{-\infty }{\rm e}^{n\xi }\,{\rm d}\xi =-\frac{1}{n}u[\alpha
(-\infty )].
\end{eqnarray*}
The limit $t\rightarrow +\infty $ is similar. Dependence of $v$ on $\alpha
(t)$ can be checked by taking the derivative with respect to $\alpha (t):$%
\[
\frac{{\rm d}v}{{\rm d}t}=-\int_{0}^{-\infty }{\rm e}^{n\xi }\,\,\beta
[\alpha (t-\xi )]\frac{\partial u[\alpha (t-\xi )]}{\partial \ln\alpha
(t-\xi )}\,\,{\rm d}\xi =-\beta [\alpha (t)]\frac{\partial v}{\partial
\ln\alpha (t)}. 
\]
In the last step we have used the equality 
\[
\beta [\alpha (t)]\frac{\partial }{\partial\ln \alpha (t)}=\beta [\alpha (s)]%
\frac{\partial }{\partial \ln\alpha (s)} 
\]
for any $t$ and $s$.

We finally observe that the equality $nv+u=0$, holding at the fixed points,
assures that $u$ and $v$ have the same behavior at criticality. In
particular, $u\sim\beta^2$ implies $v\sim \beta^2$, which is why we collect
a $\beta^2$ in front of $\tilde f$ in (\ref{tetateta'}). We conclude that $%
\tilde{f}$ is positive and depends on the running coupling constant.

Some arguments work also for the case $n=0$, if $u$ tends to zero
sufficiently fast at criticality. In that case we can write 
\[
v\,[\alpha (t)]=v[\alpha (\infty )]-\int_{t}^{\infty }\,u[\alpha (t^{\prime
})]\,\,{\rm d}t^{\prime }, 
\]
but there is no unambiguous way to fix the additive constant, so that the
sign of the function $v$ is in general not fixed.

Using (\ref{dd}) we can also write 
\begin{equation}
a^{\prime}_{{\rm IR}}-a^{\prime}_{{\rm UV}}=-\int_{-\infty }^{+\infty }{\rm d%
}t\ \beta ^{2}(t)\tilde{f}(t),  \label{bis2}
\end{equation}
since all the terms containing $\frac{{\rm d}}{{\rm d}t}$ integrate
straightforwardly to zero.

\subsection{Interpolation between the critical values}

We now study the correlator (\ref{tetateta}) at $x=y$ as well as $%
|x-y|=\infty $, which we can write in the form 
\begin{equation}
\langle\Theta (x)\ \Theta (y)\rangle=-\frac{1}{90}\frac{1}{(4\pi )^{2}}\Box
^{2}\left[ a_{{\rm UV}}^{\prime }\delta (x-y)-\frac{1}{2\pi ^{2}}\frac{\beta
^{2}(t)\tilde{f}(t)}{|x-y|^{4}}-a_{{\rm IR}}^{\prime }\frac{1}{|x-y|^{8}}%
\delta \left( \frac{x-y}{|x-y|^{2}}\right) \right] ,  \label{correction}
\end{equation}
Let us first discuss the singularities at $x=y$. It is easy to see that this
correlator, in particular the central non-local term, is well defined as a
distribution. For the study of the divergent part, we can ignore the overall 
$\Box ^{2}$. We have 
\[
\int {\rm d}^{4}x~u(x-y)\frac{\beta ^{2}(t)\tilde{f}(t)}{|x-y|^{4}}<\infty 
\]
for any regular bounded test function $u$. This means that the perturbative
divergences sum up and disappear once the cut-off is removed. This situation
is common when dealing with anomalies and, in general, evanescent operators 
\cite{collins}. Certainly we can have information from the perturbative
divergences before removing the cut-off \cite{hathrell}, but the final
correlator is convergent.

Note the ``$\delta $-function at infinity'', that we include formally in (%
\ref{correction}), required by conformal invariance. When the theory is
conformal the middle term vanishes and $a_{{\rm UV}}^{\prime }=a_{{\rm IR}%
}^{\prime }=a_{*}^{\prime }$, so that 
\begin{equation}
\langle\Theta (x)\ \Theta (y)\rangle=-\frac{1}{90}\frac{1}{(4\pi )^{2}}%
a_{*}^{\prime }\Box ^{2}\left[ \delta (x-y)-\frac{1}{|x-y|^{8}}\delta \left( 
\frac{x-y}{|x-y|^{2}}\right) \right] ,  \label{cri}
\end{equation}
which is indeed conformal-invariant. Formula (\ref{correction}) expresses
that $a_{{\rm UV}}^{\prime }$ and $a_{{\rm IR}}^{\prime }$ are the small-
and large-distance limits of $a^{\prime}(r)$, respectively, and the
non-local term interpolates between the two. For a running theory we have
necessarily $a_{{\rm UV}}^{\prime }\neq a_{{\rm IR}}^{\prime }$ (actually $%
a_{{\rm UV}}^{\prime }>a_{{\rm IR}}^{\prime }$). We would like to describe
this interpolation in more detail.

By performing a rescaling $\mu \rightarrow \lambda \mu $ we can prove the
following limits: 
\begin{eqnarray}
\lim_{\lambda \rightarrow 0}\frac{1}{2\pi ^{2}}\frac{\beta ^{2}(t+\ln
\lambda )\tilde{f}(t+\ln \lambda )}{|x-y|^{4}} &=&(a_{{\rm UV}}^{\prime }-a_{%
{\rm IR}}^{\prime })\frac{1}{|x-y|^{8}}\delta \left( \frac{x-y}{|x-y|^{2}}%
\right) ,  \label{dis} \\
\lim_{\lambda \rightarrow \infty }\frac{1}{2\pi ^{2}}\frac{\beta ^{2}(t+\ln
\lambda )\tilde{f}(t+\ln \lambda )}{|x-y|^{4}} &=&(a_{{\rm UV}}^{\prime }-a_{%
{\rm IR}}^{\prime })\delta (x-y).  \label{dis2}
\end{eqnarray}

Again, these formulas have to be meant in the sense of distributions, so
their proofs are worked out by means of a test function. We have 
\[
\frac{1}{2\pi ^{2}}\int {\rm d}^{4}x\ u(|x-y|)\frac{\beta ^{2}(t+\ln \lambda
)\tilde{f}(t+\ln \lambda )}{|x-y|^{4}}=\int_{-\infty }^{+\infty }{\rm d}t\
\beta ^{2}(t)\tilde{f}(t)u(|x-y|/\lambda )\rightarrow (a_{{\rm UV}}^{\prime
}-a_{{\rm IR}}^{\prime })u(1/\lambda ) 
\]
for $\lambda \rightarrow \infty ,0$. In the case $\lambda \rightarrow 0$,
formula (\ref{dis}) is recovered. On the other hand, in the limit $\lambda
\rightarrow \infty $ the result $(a_{{\rm UV}}-a_{{\rm IR}})u(\infty )$ is
also in agreement with (\ref{dis2}).

Using (\ref{dis}) and (\ref{dis2}) we see that, in the UV limit, (\ref
{correction}) tends to formula (\ref{cri}) with $a_{*}^{\prime }=a_{{\rm UV}%
}^{\prime }$. Similarly (\ref{dis}) shows that, in the IR limit, (\ref
{correction}) tends also to formula (\ref{cri}) with $a_{*}^{\prime }=a_{%
{\rm IR}}^{\prime }$. We have therefore proved that the correlator $%
\langle\Theta \Theta \rangle$ interpolates between the UV\ and IR\ values of
the coefficient of the term $\Box R$ in the trace anomaly operator equation.

\subsection{Scheme independence in the presence of scalar fields}

It is important that the function $f$ depends only on the running coupling $%
\alpha (t),$ i.e. that it does not depend explicitly on $\alpha (\mu )$, as
a consequence of the Callan--Symanzik equations and the finiteness of the
stress--energy tensor. However, it is well known that in the presence of
scalar fields $\varphi $ the stress--energy tensor is not truly finite \cite
{zinnjustin,ccfis}. It mixes with the operator $\left( \partial _{\mu
}\partial _{\nu }-\eta _{\mu \nu }\Box \right) \varphi ^{2}$ and therefore a
proper definition of a physical (i.e. scheme-independent) function depending
on the coupling constant at a single energy scale is more subtle. A\
function of this type was called {\it central function} in ref. \cite{ccfis}%
. In the example of the previous section the central function was $\beta
^{2}f$:\ it is physical since it is defined by a physical correlator ($f$
alone, instead, is scheme-dependent, since $\beta $ is). In general, the
two-point function of an operator ${\cal O}$ (with canonical dimension $d$
and, for simplicity, not mixing with other operators) can be written in the
form 
\begin{equation}
\langle{\cal O}(x)\,{\cal O}(y)\rangle=\frac{Z^{2}[\alpha (t),\alpha (\mu
),s]A[\alpha (t),s]}{|x-y|^{2d}}=\frac{B[\alpha (t),\alpha (\mu )]}{%
|x-y|^{2d}}.  \label{fr}
\end{equation}
In order to be as explicit as possible, we use a (temporary) heavy notation.
The ``variable'' $s$ refers to scheme dependence. The renormalization
constant $Z$ depends in general on the values of the coupling constant at
two different energy scales: 
\[
Z=\exp \left( \int_{\alpha (\mu )}^{\alpha (t)}\frac{\gamma (\alpha ){\rm d}%
\alpha }{\beta (\alpha )}\right) 
\]
and on the subtraction scheme, since both $\beta $ and $\gamma $ do. Formula
(\ref{fr}) contains no central function: each function appearing there is
either scheme-dependent or depends on the couplings at two different energy
scales (which prevents from defining univocal critical limits).

If the operator ${\cal O}$ is a conserved current, then $Z=1$ and $%
B=B[\alpha (t)]$ has the desired properties. In passing, we note that $B$ is
scheme-independent although the running coupling constant $\alpha (t)$
depends on the scheme, $\alpha (t)=\alpha (t,s)$. The reason is that the
function $B$, {\it as a function of }$\alpha ,$ is also scheme-dependent, $%
B=B(\alpha ,s)$, while it is scheme-independent {\it as a function of }$t$:
in $B(t)=B[\alpha (t,s),s]$. In the physical correlator the two scheme
dependences cancel each other.

If the operator ${\cal O}$ is the stress-energy tensor and scalar fields are
present, then we need an additional effort to identify the desired central
function. As in ref. \cite{ccfis}, the matrix of renormalization constants
for the couple $({\cal O}_{1},{\cal O}_{2})\equiv (\Theta ,\square \phi
^{2}) $ of mixing operators is triangular, 
\begin{equation}
Z_{ij}=\left( 
\begin{array}{ll}
1 & \xi \\ 
0 & \zeta
\end{array}
\right) =Z_{ij}[\alpha (t),\alpha (\mu ),s],  \label{q}
\end{equation}
$\zeta $ being the renormalization constant of the mass operator $\phi ^{2}$%
. We have 
\begin{equation}
\langle{\cal O}_{i}(x)\,{\cal O}_{j}(0)\rangle={\frac{1}{|x|^{4}}}%
Z_{ik}[\alpha (t),\alpha (\mu ),s]Z_{jl}[\alpha (t),\alpha (\mu
),s]F_{kl}[\alpha (t),s]\equiv {\frac{1}{|x|^{4}}}P_{ij}[\alpha (t),\alpha
(\mu )],  \label{w}
\end{equation}
where we have exhibited all dependences as in (\ref{fr}). Writing 
\begin{equation}
P_{ij}[\alpha (t),\alpha (\mu )]\equiv \left( 
\begin{array}{ll}
p & r \\ 
r & q
\end{array}
\right) ,\qquad F_{kl}[\alpha (t),s]\equiv \left( 
\begin{array}{ll}
k & h \\ 
h & g
\end{array}
\right)  \label{matrices}
\end{equation}
and combining (\ref{q}), (\ref{w}) and (\ref{matrices}) we find 
\begin{equation}
\frac{\det P}{q}=p-\frac{r^{2}}{q}=\frac{\det F}{g}=k-\frac{h^{2}}{g}\equiv
\beta ^{2}[\alpha (t)]f[\alpha (t)]\geq 0.  \label{invariant}
\end{equation}
Now, $p-r^{2}/q$ is manifestly scheme-independent, while $k-h^{2}/g$ does
not depend on $\alpha (\mu )$. The equality of the two expressions allows us
to conclude that both expressions are scheme-independent and functions of
the running coupling $\alpha (t)$. This defines the desired central function
when scalar fields are present, which we write as $\beta ^{2}[\alpha
(t)]f[\alpha (t)]$. It has to be inserted into formulas (\ref{resu}), (\ref
{dir}), (\ref{bis2}), etc., to give the general formula of the $a^{\prime}$%
-function. Formula (\ref{invariant}) gives the only invariant of the
similarity transformation $P=ZFZ^{t}$, with $P$ and $F$ symmetric and $Z$
triangular of the form (\ref{q}).

We have factorized out a $\beta ^{2}$: indeed, $p\sim \langle\Theta \,
\Theta \rangle$ is proportional to $\beta ^{2}$, while $r\sim \langle\Theta
\,\Box \varphi ^{2}\rangle$ is proportional to $\beta $. Finally, the
denominator is regular since the function $g$ is regular at criticality. To
see this, one observes that the factor $\zeta $ in the correlator $%
\langle\Box \varphi ^{2}\, \Box \varphi ^{2}\rangle=q/|x|^{8}=\zeta
^{2}g/|x|^{8}$ takes care of the eventual anomalous dimension $\gamma $ of
the operator $\varphi ^{2}$ at criticality ($\zeta \sim 1/|x|^{2\gamma }$),
so that $g$ remains non-vanishing (and positive by applying reflection
positivity to this correlator).

Finally, reflection positivity of $\langle{\cal O}_{i}\, {\cal O}_{j}\rangle$
assures that the matrix $P$ is positive-definite. In particular, $\det P$ is
positive. Since $q$ is also positive, $f\geq 0$ follows.

The existence of the (unique) invariant (\ref{invariant}) for the similarity
transformation that relates $P$ and $F$ by the matrix $Z$ of (\ref{q}) is an
important fact; it assures that scalar fields are included in the treatment.
By comparing the calculations done in \cite{hathrell2} and \cite{hathrell}
one can appreciate the additional amount of effort required by scalar fields.

Our discussion about scheme dependence is an introduction to the notion of
``covariance'' that we will formulate later on in this context (section 3.5).

\section{The critical values of the quantity $a^{\prime }$: normalization,
physical meaning and its relationship with $a$}

It is a consequence of our arguments that even if the quantity $a^{\prime}$
is to some extent undetermined, its RG group flow is uniquely determined.
For example, the difference $a^{\prime}_{{\rm UV}}-a^{\prime}_{{\rm IR}}$ is
the invariant area of the graph of the beta function and the derivative of $%
a^{\prime}$ is expressed in terms of a physical correlator, $\langle\Theta
(x)\ \Theta (y)\rangle$. Therefore, the ambiguity of $a^{\prime}$ can be at
most an additive constant, so that once we normalize it at a reference
energy scale (for example in the UV\ limit) then it is fixed at any energy
scale. Perturbative calculations allow us to arrive at the same conclusion 
\cite{hathrell}.

Nevertheless, there is a preferred normalization choice for $a^{\prime }$.
Indeed, it turns out that the quantities $a$ and $a^{\prime }$ have various
properties in common. For example, they are both two-loop-uncorrected, while 
$c$ is two-loop-corrected. The relation between $c$ and $a$ that is
sometimes advocated in the literature \cite{birreldavies}, instead, is an
artefact of the regularization technique. The radiative corrections of both $%
a$ and $a^{\prime }$ begin at three loops. Actually, they coincide (see
sect. 3.2).

\subsection{The Riegert action}

The four-dimensional analogue of the Polyakov action $S_{{\rm P}}$ in two
dimensions \cite{polyakov}, 
\[
S_{{\rm P}}=-{\frac{c}{48\pi }}\int {\rm d}^{2}x\sqrt{-g_{x}}\int {\rm d}%
^{2} y\sqrt{-g_{y}}~R_{x}\square _{(x,y)}^{-1}R_{y}, 
\]
has been worked out and studied in several papers. The more ancient article
containing the complete non-local action is, to our knowledge, the one by
Riegert \cite{riegert}. The local action for the conformal factor was found
also by Fradkin and Tseytlin in ref. \cite{fradkin}. 
Buchbinder {\it at al.} were able to treat also
the case with non-vanishing torsion.
More recent studies are
those by Cappelli and Coste \cite{coste}, 
Antoniadis and Mottola \cite{anto}, 
and others \cite{mazur,others}. We take the expression of the action from 
\cite{anto}, which is more symmetric than the one given by Riegert (the
difference is a conformal-invariant term). One has to integrate the equation 
\[
\Theta =-2\frac{1}{\sqrt{-g}}g_{\mu \nu }\frac{\delta S_{{\rm R}}}{\delta
g_{\mu \nu }}. 
\]
One can stay at criticality and use expression (\ref{critica}) for $\Theta $%
. The result is, in our notation, 
\begin{eqnarray*}
S_{{\rm R}} &=&-\frac{1}{160(4\pi )^{2}}\frac{1}{a_{*}}\int {\rm d}^{4}x%
\sqrt{-g_{x}}\int {\rm d}^{4}x\sqrt{-g_{y}}\left[ c_{*}W^{2}-{\frac{1}{3}}
a_{*}\left({\rm G}-\frac{2}{3}\Box R\right)\right] _{x} \\
&&\left[ 2\Box ^{2}+4R^{\mu \nu }\nabla _{\mu }\nabla _{\nu }-\frac{4}{3}%
R\Box +\frac{2}{3}(\nabla ^{\mu }R)\nabla _{\mu }\right] _{(x,y)}^{-1}\left[
c_{*}W^{2}-{\frac{1}{3}}a_{*}\left({\rm G}-\frac{2}{3}\Box R\right)\right]
_{y} \\
&&+\frac{a_{*}-a_{*}^{\prime }}{6480(4\pi )^{2}}\int {\rm d}^{4}x\sqrt{-g}%
R^{2}.
\end{eqnarray*}
We stress that the Riegert action, obtained by integrating the trace
anomaly, is not the complete induced action: it misses the
conformal-invariant terms (local as well as non-local). Only in two
dimensions is a conformally invariant term gauge-equivalent to zero.

We rederive $S_{{\rm R}}$ in the particular case of a metric of the form $%
g_{\mu \nu }={\rm e}^{2\phi }\eta _{\mu \nu },$ which is sufficient for our
purposes.

Let us start from formula (\ref{critica}). The induced action $S_{{\rm R}%
}[\phi ]$ for the conformal factor $\phi $ is the solution of the equation 
\begin{equation}
\Theta =-{\rm e}^{-4\phi }\frac{\delta S_{{\rm R}}[\phi ]}{\delta \phi }.
\label{equ}
\end{equation}
One observes that the combination ${\rm G}-\frac{2}{3}\Box R$ is very simple
(see \cite{anto}): 
\[
{\rm G}-\frac{2}{3}\Box R=4{\rm e}^{-4\phi }\Box ^{2}\phi . 
\]
Then formula (\ref{critica}) gives, in the case of a conformally flat
metric, 
\[
\Theta =\frac{1}{90(4\pi )^{2}}\left[ a_{*}{\rm e}^{-4\phi }\Box ^{2}\phi +%
\frac{1}{6}(a_{*}-a_{*}^{\prime })\Box R\right] . 
\]
It is not necessary to write $\Box R$ explicitly, since it can be integrated
using 
\[
\Box R=-\frac{1}{6\sqrt{-g}}g_{\mu \nu }^{{}}\frac{\delta }{\delta g_{\mu
\nu }^{{}}}\int {\rm d}^{4}x\sqrt{-g}R^{2}. 
\]
The solution of eq. (\ref{equ}) is then straightforward. The result is 
\begin{equation}
S_{{\rm R}}[\phi ]=-\frac{1}{180}\frac{1}{(4\pi )^{2}}\int {\rm d}%
^{4}x\left\{ a_{*}(\Box \phi )^{2}-(a_{*}-a_{*}^{\prime })\left[ \Box \phi
+(\partial _{\mu }\phi )^{2}\right] ^{2}\right\} .  \label{lor}
\end{equation}

From this action, we immediately recover (\ref{aprimo}) with the correct
sign (remember that (\ref{lor}) is written in the Lorentzian framework).
Turning the exponential factor ${\rm e}^{iS[\phi ]}$ to the Euclidean
framework, one has ${\rm e}^{-S_{{\rm E}}[\phi ]},$ in terms of the
Euclidean action 
\[
S_{{\rm E}}=\frac{1}{180}\frac{1}{(4\pi )^{2}}\int {\rm d}^{4}x\left\{
a_{*}(\Box \phi )^{2}-(a_{*}-a_{*}^{\prime })\left[ \Box \phi +(\partial
_{\mu }\phi )^{2}\right] ^{2}\right\} . 
\]
For $a^{\prime }=a$ this action is free, which means that the three- and
four-point functions of $\Theta $ are zero at criticality (as well as all
the correlators $\langle \Theta(x_1)\cdots \Theta(x_n)\rangle$, $n>2$) and
that the two-point function $\langle \Theta (x)~\Theta (0)\rangle $ equals $-%
\frac{1}{90(4\pi) ^{2}}a\Box ^{2}\delta (x)$.

With the identification $a^{\prime }=a$, we can suppress the primes in
formula (\ref{resu}) and finally state the $a$-theorem.

\vskip .3truecm

{\bf $a$-theorem}.

i) $a$ {\it is non-negative.}

ii) {\it The total RG flow of $a$ is non-negative and equal to the invariant
area of the beta function:} 
\begin{equation}
a_{{\rm UV}}-a_{{\rm IR}}=-\int_{\alpha _{{\rm UV}}}^{\alpha _{{\rm IR}}}%
\frac{{\rm d}\alpha }{\alpha }\ \beta (\alpha )f(\alpha )\geq 0.
\label{resu2}
\end{equation}

\vskip .2truecm

This prediction will be checked in section 3.3 to the fourth-loop order in
perturbation theory, for QCD in the conformal window around the asymptotic
freedom point $N_{f}={\frac{11}{2}}N_{c}$, as well as supersymmetric QCD,
QED and the $\varphi ^{4}$-theory.

In realistic UV free theories both the UV and IR fixed points are free
theories and therefore $a_{{\rm UV}}-a_{{\rm IR}}$ is a positive, integer
number. In this case we see that the invariant area of the beta function is
quantized, with a unit cell for each massless degree of freedom that
disappears along the renormalization group flow. Furthermore, the $a^{\prime
}$-function of the previous section correctly interpolates between the
critical values $a_{{\rm UV}}\geq a_{{\rm IR}}\geq 0,$ so that at each
intermediate energy the flow of $a$ equals the invariant area of the graph
spanned by the beta function up to that energy.

Finally, the relation $a_{*}=a_{*}^{\prime }$ has other interesting
implications when the conformal factor is quantized \cite{anto,mazur}.

\subsection{Discussion of the positive-definiteness of the Riegert action and%
\newline
its consequences}

In this section I present a discussion about unitarity and the
positive-definiteness of the Riegert action 
and give an argument for the $a$-theorem.


A consequence of unitarity is that

{\it if the classical action }$S_{cl}[\varphi ]${\it \ is positive-definite
(in the Euclidean framework), then the quantum action is positive-definite.}

We refer in particular to the functional generator $\Gamma [\varphi ]$ of
connected one-particle irreducible diagrams. Note that positive-definiteness
does not imply the existence of a minimum. Indeed, there are examples where
the quantum action does not have a minimum. Positivity of the classical
action assures that the functional integral is well defined, while
positivity of the functional generator $\Gamma $ is the statement that the
resulting theory makes physical sense. We assume the the additive constants
of $S_{cl}$ and $\Gamma$ are adjusted so that positive definiteness is
equivalent to boundedness from below.

In stating the above property we are thinking of bosonic actions (classical
and quantum). Fermions can be included in the classical action with no
problem. On the other hand, we are mostly interested in induced actions for
bosonic fields and sources.

The above statement would be trivial in the absence of divergences, but the
regularization cuts off certain frequencies and therefore violates
unitarity. The statement is therefore false in the regularized theory.
Renormalization can be seen as the process of restoring positivity by
compensating the undesirable effects of logarithms.

For example, the quadratic part of the induced action of fermions in an
external electromagnetic field reads in momentum space 
\begin{equation}
-\frac{\beta _{1}}{32\pi}\int {\frac{{\rm d}^{4}p}{(2\pi)^4}}
\left|F_{\mu\nu}(p)\right|^2\ln p^{2}/\Lambda ^{2},  \label{em1}
\end{equation}
where $\beta _{1}$ is the one-loop coefficient of the beta function ($\beta
=\beta _{1}\alpha +{\cal O}(\alpha ^{2})$) and $\Lambda $ a cut-off. This
expression is either positive or negative, depending on $\beta _{1},$ $%
\Lambda $ and the evaluation of the integral. If, however, the
electromagnetic field is dynamical, there will be an additional contribution 
\begin{equation}
\frac{1}{16\pi\alpha (\Lambda )}\int {\frac{{\rm d}^{4}p}{(2\pi)^4}}%
\left|F_{\mu \nu }(p)\right|^{2},  \label{em2}
\end{equation}
which removes the divergence and restores positivity. Indeed, $\alpha
(\Lambda )$ is defined in such a way that the sum 
\begin{equation}
{\frac{1}{16\pi}}\int {\frac{{\rm d}^{4}p}{(2\pi)^4}}\left|F_{\mu \nu
}(p)\right|^{2} \left({\frac{1}{\alpha(\Lambda)}}-{\frac{\beta_1}{2}}\ln
p^2/\Lambda^2\right)  \label{temp}
\end{equation}
is independent of $\Lambda $. Now, the first term is always positive, while
the negative contribution coming from the second term is originated by the
region $|p|<\Lambda $ if $\beta _{1}<0$ and $|p|>\Lambda $ if $\beta _{1}>0.$
Either region is arbitrarily small in a suitable limit for $\Lambda $ ($%
\Lambda \rightarrow 0$ in the first case, which is the case of asymptotic
freedom, and $\Lambda \rightarrow \infty $ in the second case, which is the
case of IR\ freedom) and therefore every negative contribution is
reabsorbed. Correctly, in these limits $\alpha (\Lambda )\rightarrow 0$ to
compensate for the infinitely large negative contributions coming from the
second term of (\ref{temp}). 
Here there is a Landau pole, so that complete positivity is not
restored order-by-order in perturbation theory, but just kept far from
the perturbative regime. Positivity must be fully recovered, 
however, in the
complete theory, which has to be unitary. 

Equivalently, one defines a running coupling $\alpha (|p|)$ and writes the
action (\ref{temp}) as 
\[
{\frac{1}{16\pi }}\int {\frac{{\rm d}^{4}p}{(2\pi )^{4}}}\left| F_{\mu \nu
}(p)\right| ^{2}{\frac{1}{\alpha (|p|)}}.
\]
Then the statement is that the running coupling constant is everywhere
positive, once it is positive at a given energy. This is clearly
visible in the conformal window, where the running 
coupling constant is indeed positive at all energies. Positivity at a given
energy is assured by the physical normalization. 

We remark that it is crucial to {\it have} a parameter to reabsorb the
violations of positivity. For example, if we just subtract the divergent
part of (\ref{em1}), but we do not do it with the help of a new (running)
parameter we get an expression like 
\begin{equation}
-\frac{\beta _{1}}{32\pi }\int {\frac{{\rm d}^{4}p}{(2\pi )^{4}}}\left|
F_{\mu \nu }(p)\right| ^{2}\ln p^{2}/\mu ^{2},  \label{em3}
\end{equation}
with $\mu $ arbitrary and finite. This expression is convergent, but has no
chance to be positive. Renormalization is not just a pure subtraction of
divergences, but, more deeply, it is the unique way to restore positivity.

In summary, the violations of positivity are parametrized by local terms and
can be reabsorbed by appropriate local counterterms, multiplied by physical
parameters that can eventually run in order to reabsorb the violations of
positivity. If this goal cannot be achieved then the theory has negatively
normed states and unitarity is violated.

These observations, we think, suggest an instructive way to look at
renormalization, that we have not found emphasized in the existing literature%
\footnotemark 
\footnotetext{
We plan to elaborate further on this approach elsewhere.}.

\subsubsection*{Induced action for the external conformal factor}

We now make a further step and consider induced actions for external sources 
$\phi _{{\rm ext}}$. If we could ignore the problems associated with
divergences we could state that

{\it if the classical action $S_{cl}[\varphi,\phi]$ is positive definite in
the full space of fields $\varphi$ and sources $\phi$, then the quantum
action is positive definite in the full space of fields and sources.}

In general, external sources are not such that the classical action $%
S_{cl}[\varphi,\phi]$ is positive definite in the full space of fields and
sources, but the (minimal) coupling to (external) gravity does satisfy this
requirement. 

The above statement is again spoiled by divergences or effects related to
divergences (like the anomalies). Moreover, since an induced action for
external fields is not equipped with the appropriate parameters under which
the violations of positivity can be reabsorbed, it might have an indefinite
sign.

For example, the Polyakov action $S_{{\rm P}}$ in two dimensions is negative
definite and convergent. The Riegert action in four dimensions, as we are
going to discuss, is positive and convergent. The quantities $c$, $a$, $%
a^{\prime }$ and $h$ can be thought of as the matter contributions to the
beta functions of higher-derivative quantum gravity and there is no reason
why they should have a definite sign.

What we can expect nevertheless, is that if the induced action is convergent
and positive at some energy scale, than it is positive at all energy scales.
This statement well applies to our case.


In order to have a convergent induced effective action, one should consider
sources coupled to evanescent operators \cite{collins}. The Riegert
effective action, moreover, is convergent (notwithstanding the inverted $%
\Box ^{2}$-operator), up to conformal-invariant terms (we recall that the
Riegert action is defined up to such terms) and total derivatives. Indeed,
the divergence due to the inverted $\Box ^{2}$-operator is proportional to 
\[
\left( \int {\rm d}^{4}x\,\sqrt{g}\left[ c_{*}W^{2}-{\frac{1}{3}}a_{*}\left(
G-{\frac{2}{3}}\Box R\right) \right] \right) ^{2}
\]
and therefore harmless. Similarly in two dimensions the divergence is
proportional to a total derivative: 
\[
\left( \int {\rm d}^{2}x\,\sqrt{g}R\right) ^{2}.
\]
Once we specialize to the conformal factor $\phi $, coupled to the
evanescent operator $\Theta $, convergence is more apparent. This can be
seen also off-criticality, where the quadratic part of the effective action
is read from correlator (\ref{correction}). In momentum space we have
(dropping the contribution from infinity) 
\[
{\frac{1}{180(4\pi )^{2}}}\int {\frac{{\rm d}^{4}p}{(2\pi )^{4}}}%
(p^{2})^{2}\left| \tilde{\phi}(p)\right| ^{2}\left[ a_{{\rm IR}}^{\prime }+{%
\frac{1}{2\pi ^{2}}}\int {\rm d}^{4}x\,\left( 1-\cos (p\cdot x)\right) {%
\frac{\beta ^{2}(t)\tilde{f}(t)}{|x|^{4}}}\right] ,
\]
which is convergent, and positive as long as $a_{{\rm IR}}^{\prime }$ is.
The expression between brackets is equal to $a_{{\rm UV}}^{\prime }$ in the
limit $|p|\rightarrow \infty $ and to $a_{{\rm IR}}^{\prime }$ in the limit $%
|p|\rightarrow 0$ (but note that this expression does not interpolate
monotonically between the two values; for this purpose one should use the
function $a^{\prime }(t)$ constructed in section 2.1). As we see, there
are cases where it is
relatively easy to get positivity at all energies and this aspect of the
problem is controlled by the local terms.

Convergence is an important property in the context of our discussion,
because there is no local counterterm that can cure the first term, $\int
(\Box \phi )^{2}$, in the Riegert action $S_{{\rm R}}$. The second term,
instead, $\int R^{2}$, should be cured by the $a^{\prime }$-ambiguity
itself. We conclude that the total action should be positive-definite
throughout the RG\ flow if it is at some intermediate energy. In the next
subsections we show that this statement is equivalent to the 
full $a$-theorem.

\subsubsection*{Point {\rm (i)} of the $a$-theorem
}

The total induced gravitational action contains three types of terms. The
conformally invariant terms, convergent or divergent as they might be, are
not visible in the Riegert action, which contains the other two types of
terms. The first one is $\int (\Box \phi)^{2}$, with coefficient $a$, which
we discuss here. The second term, $\int \sqrt{g}R^2$, is discussed in the
next subsection.

There is no arbitrariness that can restore the eventual positivity violation
in the term $a\int (\Box \phi )^{2}$, as we have already remarked. Our
positivity arguments, i.e. that the action should be positive definite
throughout the RG flow once it is positive definite at a reference energy,
imply that $a$ be positive also in the interacting fixed point, since
certainly $a$ is positive in the free field limit. However, it is puzzling
to have an induced action like $-c\int (\partial \phi )^{2}$ in two
dimensions, which is always negative definite. There is no contraddiction
with our statement, actually, since one might say that it does not apply to
this case, because there is no reference energy at which the induced action
is positive definite. Even better, we can observe that our statement implies
also that if the induced action is negative definite, or indefinite, at some
reference energy, then there cannot be any energy at which it is positive
definite, which is true also in two dimensions.

\subsubsection*{Point {\rm (ii)} of the $a$-theorem}

The term $\int \sqrt{g}R^2$ is not affected by divergences either. It is a
well-known fact that there cannot be any $R^2$-term in the trace anomaly at
criticality (the absence of divergences off-criticality was discussed in
section 2.2). Now, {\it there is} an arbitrary parameter associated with
this term, precisely $a^{\prime}$, and this should suffice to assure
positivity, finally explaining what the $a^{\prime}$-ambiguity is there for.
Moreover, convergence implies that $a^{\prime}$ is not a running parameter,
but just an additive constant, in agreement with the knowledge gained in the
previous sections.

We conclude that the term $\int \sqrt{g}R^2$ is positive-definite throughout
the renormalization group flow, once its coupling constant $a^{\prime}-a$ is
normalized to be positive at a given energy (we can choose one of the two
fixed points). This observation is sufficient to prove point (ii) of the $a$%
-theorem, as we now show.




Now, the term proportional to $\int \sqrt{g}R^2$ is bounded from below at
criticality if 
\begin{equation}
a_{*}^{\prime }\geq a_{*}.\qquad  \label{twocond}
\end{equation}
This condition has to hold throughout the renormalization group flow, in
particular 
\[
a^{\prime}_{{\rm UV}}\geq a_{{\rm UV}}~~~~~~~\Leftrightarrow
~~~~~~~a^{\prime}_{{\rm IR}}\geq a_{{\rm IR}}. 
\]
Now, we know that 
\[
a_{{\rm UV}}^{\prime }\geq a_{{\rm IR}}^{\prime }. 
\]
Let us fix $a^{\prime }$ by demanding that $a$ and $a^{\prime }$ coincide in
the UV, $a_{{\rm UV}}^{\prime }=a_{{\rm UV}}$. Then we have, combining the
various inequalities derived so far: 
\[
a_{{\rm UV}}=a_{{\rm UV}}^{\prime }\geq a_{{\rm IR}}^{\prime }\geq a_{{\rm IR%
}}, 
\]
wherefrom the claimed inequality $a_{{\rm UV}}\geq a_{{\rm IR}}$ follows.

Now, let us tentatively suppose that with the normalization $a_{{\rm UV}%
}^{\prime }=a_{{\rm UV}}$ we have the strict inequality $a_{{\rm IR}%
}^{\prime }>a_{{\rm IR}}$. We prove that this is absurd and conclude that $%
a_{{\rm IR}}^{\prime }=a_{{\rm IR}}$.

We can do this by changing the normalization of $a^{\prime }$ with the shift 
$a^{\prime }\rightarrow a^{\prime ~{\rm new}}=a^{\prime }-a_{{\rm IR}%
}^{\prime }+a_{{\rm IR}},$ so that $a_{{\rm IR}}^{\prime ~{\rm new}}=a_{{\rm %
IR}}$. We have $a_{{\rm UV}}^{\prime }\rightarrow $ $a_{{\rm UV}}^{\prime ~%
{\rm new}}=a_{{\rm UV}}^{\prime }-a_{{\rm IR}}^{\prime }+a_{{\rm IR}}$ and
therefore $a_{{\rm UV}}^{\prime }$ no longer satisfies the inequality (\ref
{twocond}): $a_{{\rm UV}}^{\prime ~{\rm new}}<a_{{\rm UV}}$. This is a
contradiction. We conclude that 
\begin{equation}
a_{{\rm UV}}^{\prime }=a_{{\rm UV}}\qquad \Leftrightarrow \qquad a_{{\rm IR}%
}^{\prime }=a_{{\rm IR}}.  \label{unita}
\end{equation}

In conclusion, the total RG\ flows of $a$ and $a^{\prime }$ are equal and
given by formula (\ref{resu2}). The identification $a^{\prime}=a$ is
consistent and the difference $a_{{\rm UV}}-a_{{\rm IR}}$ is equal to the
area of the graph of the beta function. The interplay between unitarity and
renormalization is able to turn a simple set of inequalities into a precise
non-perturbative formula.

Another way to state our result is that the coupling constant for the term $%
R^{2}\sqrt{g}$, which is $a^{\prime }-a$, is non-renormalized. This is not
surprising, in the end, since the running behaviour of a coupling constant
is due to divergent contributions, but the $a^{\prime }$-ambiguity is fully
finite.

Note that the result (\ref{unita}) would follow even if positivity implied $%
a^{\prime}_*\leq a_*$ instead of (\ref{twocond}).

\subsection{Perturbative checks}

In this section we check our predictions to the fourth loop order in
perturbation theory around the free fixed point. The strategy for computing
higher-loop corrections to the trace anomaly was formulated by Brown and
Collins \cite{brown}, applied by Hathrell \cite{hathrell,hathrell2} and
Freeman \cite{freeman} to the third-loop order, and extended by Jack and
Osborn to the fourth-loop order (and in several other directions) \cite{jack}%
.

We begin with the third loop analysis and use mostly the results of refs. 
\cite{hathrell,hathrell2}, since, to our knowledge, the papers by Hathrell
are the only ones in which the term $\Box R$ is treated explicitly. The
paper by Freeman does not calculate $a^{\prime}$, but it contains enough
information to derive it, once Hathrell's formulas are used. Moreover, the
Hathrell--Freeman results are easily extended to a general third-loop
expression that can be directly applied, in particular, to the QCD conformal
window, in the neighbourhood of the asymptotic-freedom point. Such a formula
shows perfect agreement with our prediction, i.e. that the total RG flows of 
$a$ and $a^{\prime}$ coincide and are equal to the invariant area of the
graph of the beta function.

For the purposes of this section, there is no difference between tilded and
untilded quantities of section 1. Indeed, we are just interested in
comparing critical values and flows of critical values.

In massless QED we have \cite{hathrell} 
\begin{eqnarray*}
\tilde{c}=18+\frac{70}{3}\frac{\alpha }{4\pi }+{\cal O}(\alpha ^{2}),\quad &&%
\tilde{a}=73-180\left( \frac{\alpha }{4\pi }\right) ^{2}+{\cal O}(\alpha
^{3}),\quad \\
\tilde{a^{\prime}}=a^{\prime}_{*}-180\left( \frac{\alpha }{4\pi }\right)
^{2}+{\cal O}(\alpha ^{3}),\quad &&f_{*}=45.
\end{eqnarray*}
These values are independent of the subtraction scheme. While $\tilde{c}$
required an independent calculation, the perturbative corrections of $\tilde{%
a}$ and $\tilde{a^{\prime}}$ were computed from each other and turn out to
be equal. This coincidence is already what we need, nevertheless the
perturbative check is not exhausted by this observation, since the above
flows turn out to have the wrong signs. The explanation of this fact will
emerge from the final formula. For the moment, we keep this observation in
mind. 

Similarly, in pure Yang--Mills theory with gauge group $G$ we have \cite
{freeman} 
\begin{eqnarray*}
\tilde{c}=4\dim G\ \left[ 3-\frac{20}{3}\frac{\alpha }{4\pi }C(G)+{\cal O}%
(\alpha ^{2})\right] ,&&\tilde{a}=2\dim G\ \left[ 31+255\left( \frac{\alpha 
}{4\pi }\right) ^{2}C(G)^{2}+{\cal O}(\alpha ^{3})\right] , \\
\tilde{a^{\prime}}=a^{\prime}_{*}+510\dim G\ \left( \frac{\alpha }{4\pi }%
\right) ^{2}C(G)^{2}+{\cal O}(\alpha ^{3}),&&f_{*}=45\dim G.
\end{eqnarray*}
In \cite{freeman} $a^{\prime}$ is not calculated explicitly, but the formula
for the function $h(\alpha)$ is given. We have used the techniques of
Hathrell to derive $a^{\prime}$ from $h(\alpha)$. It is clear that the two
terms $\Box R$ and $R^2$ in $\Theta$ are related and one can indeed work out
the precise relationship as in \cite{hathrell}. We observe that, again, the
first perturbative corrections to $\tilde{a}$ and $\tilde{a^{\prime}}$
coincide, but have the wrong sign.

Going through Hathrell and Freeman's calculations, we have derived a very
simple general formula for the flows of $a$ and $a^{\prime }$ for
asymptotically free theories with a perturbative IR fixed point. We have 
\begin{equation}
a_{{\rm UV}}-a_{{\rm IR}}=a_{{\rm UV}}^{\prime }-a_{{\rm IR}}^{\prime }={%
\frac{1}{2}}f_{{\rm UV}}\beta _{2}\alpha _{{\rm IR}}^{2}+{\cal O}(\alpha _{%
{\rm IR}}^{3})={\frac{1}{2}}f_{{\rm UV}}{\frac{\beta _{1}^{2}}{\beta _{2}}}+%
{\cal O}\left( {\frac{\beta _{1}^{3}}{\beta _{2}^{3}}}\right) ,
\label{pertformu}
\end{equation}
where $\beta _{1}$ and $\beta _{2}$ are the first two coefficients of the
beta function, $\beta (\alpha )=\beta _{1}\alpha +\beta _{2}\alpha ^{2}+%
{\cal O}(\alpha ^{3})$. If the theory is IR-free, instead, the formula has
an additional minus sign. Hathrell and Freeman's results are correctly
reproduced (${\frac{1}{2}}f_{{\rm UV}}\beta _{2}\alpha _{{\rm IR}%
}^{2}\rightarrow {\frac{1}{2}}f_{{\rm UV}}\beta _{2}\alpha ^{2}$ for the
perturbative corrections in the absence of a fixed point): in QED, $\beta
_{1}={2/(3\pi )}$ and $\beta _{2}={1/(2\pi ^{2})}$; in pure Yang--Mills
theory, $\beta _{1}=-11N_{c}/(6\pi )$, $\beta _{2}=-17N_{c}^{2}/(12\pi ^{2})$%
. Hathrell does not observe the equality of the $a$- and $a^{\prime }$%
-flows, but it is relatively simple to show that this result is, to some
extent, implicit in the derivation, at least to the third-loop order. The
key formula is (5.27) of \cite{hathrell}.

At this point it is straightforward to show that our formula (\ref{resu2})
gives exactly the same result as (\ref{pertformu}), as we wanted.

Our theorem allows us to derive the above three-loop result in a
straightforward way. One just needs the one-loop value of $f$ and the
two-loop formula of the beta function. Extending the techniques of Hathrell
and Freeman to all orders in perturbation theory should allow us to prove
the equality of the fluxes of $a$ and $a^{\prime}$ in a direct way. An
effort in this sense is being undertaken.

The sign mismatches noted above have the following explanation: the second
coefficient $\beta_2$ is positive only when there {\it is} an IR fixed
point, while it is negative in pure Yang--Mills theory, where there is no
such point. The $a$-theorem does not need to hold, in general, when the IR
fixed point does not exist and several cases of this kind were indeed found
in ref. \cite{noi2}. In particular, $a$ is often negative off the conformal
window. It is nevertheless gratifying to observe that the coefficient in
question is precisely $\beta_2$, so that as soon as the IR fixed point
exists, the theorem is satisfied and when the theorem is not satisfied, this
is a signal that the IR fixed point does not exist (it might still exist, as
in QCD, after the introduction of the relevant non-perturbative effects, but
this would change also the formulas for the RG flows of $a$ and $a^{\prime}$%
, and in the end the $a$-theorem will have to be satisfied). The non-existence
of an interacting fixed point in QED was established in ref. \cite{jackiw}.

Concretely, in QCD with $N_f$ flavours and $N_c$ colours we have 
\[
a_{{\rm UV}}-a_{{\rm IR}}={\frac{44}{5}}N_c N_f \left(1-{\frac{11}{2}}{\frac{%
N_c}{N_f}} \right)^2=a^{\prime}_{{\rm UV}}-a^{\prime}_{{\rm IR}}. 
\]
For a generic gauge group $G$ and representation $R$ we take the two-loop
beta function from \cite{larin}. We have 
\[
a_{{\rm UV}}-a_{{\rm IR}}={\frac{605\,C(G)\,\dim G}{7C(G)+11 C(R)}}\left(1-{%
\frac{11}{4}}{\frac{C(G)}{T(R)}}\right)^2=a^{\prime}_{{\rm UV}}-a^{\prime}_{%
{\rm IR}}. 
\]

In supersymmetric QCD, formula (\ref{pertformu}) is in agreement with the
exact results of \cite{noi}. For concreteness, we take N=1 SQCD with group $%
G=SU(N_{c})$ and $N_{f}$ quanks and antiquarks in the fundamental
representation. The value of $f_{{\rm UV}}$ is still $45\dim G$ and $\beta
_{1}=-{\frac{1}{2\pi }}(3N_{c}-N_{f})$, $\beta _{2}=N_{c}N_{f}/(4\pi ^{2})$.
We have for $N_{f}\lesssim 3N_{c}$, 
\[
a_{{\rm UV}}-a_{{\rm IR}}={\frac{45}{2}}N_{c}N_{f}\left( 1-3{\frac{N_{c}}{%
N_{f}}}\right) ^{2}=a^{\prime}_{{\rm UV}}-a^{\prime}_{{\rm IR}}, 
\]
in agreement with the exact formula \cite{noi} 
\[
a_{{\rm UV}}-a_{{\rm IR}}={\frac{15}{2}}N_{c}N_{f}\left( 1-3{\frac{N_{c}}{%
N_{f}}}\right) ^{2}\left( 2+3{\frac{N_{c}}{N_{f}}}\right) 
\]
(recall that there is an additional factor of 360 with respect to ref. \cite
{noi} due to a change in the normalization).

As a further confirmation, we report the results for a scalar field $\varphi 
$ with a ${\frac{\lambda}{4!}}\varphi ^{4}$-interaction from \cite{hathrell2}%
: 
\begin{eqnarray*}
\tilde{c}=1-\frac{5}{36}\frac{\lambda ^{2}}{(4\pi )^{4}}+{\cal O}(\lambda
^{3}),\quad &&\tilde{a}=1+\frac{85}{288}\frac{\lambda ^{4}}{(4\pi )^{8}}+%
{\cal O}(\lambda ^{5}),\quad \\
\tilde{a^{\prime }}=a_{*}^{\prime }+\frac{85}{288}\frac{\lambda ^{4}}{(4\pi
)^{8}}+{\cal O}(\lambda ^{5})\quad &&f_{*}={\frac{5}{8(4\pi)^4}}.
\end{eqnarray*}
Ref. \cite{hathrell2} does not give the $a^{\prime }$ correction explicitly,
which was calculated using \cite{hathrell}. The first perturbative
corrections to $\tilde{a}$ and $\tilde{a^{\prime }}$ are still equal,
although they are related in a more complicated way, as a reflection of the
discussion of section 2.3. Nevertheless, to the second loop order we can
neglect the renormalization mixing between the stress tensor and $%
\Box\varphi^2$.

Our predictions agree with the results of Hathrell, but the formula is now
slightly different from (\ref{pertformu}). We have $\Theta=-{\frac{\beta}{4!}%
}\varphi^4$ and $\beta(\lambda)=\mu {\frac{{\rm d}\lambda}{{\rm d}\mu}}= {%
\frac{3}{(4\pi)^2}}\lambda^2 -{\frac{17}{3(4\pi)^4}}\lambda^3+{\cal O}%
(\lambda^4)= \beta_1\lambda^2+\beta_2\lambda^3+{\cal O}(\lambda^4)$, so that 
\[
a_{{\rm UV}}-a_{{\rm IR}}=\int^{\lambda_{{\rm UV}}}_0{\rm d}\lambda
\,\beta(\lambda) f(\lambda) =-{\frac{1}{12}}\beta_2f_{{\rm IR}}\lambda_{{\rm %
UV}}^4+{\cal O}(\lambda^5). 
\]
Note that in order to apply our formula correctly in the absence of an
interacting fixed point, we have to treat $\beta_1$ as an independent
(``small'') parameter and pretend that an UV fixed point does exist at $%
\lambda_{{\rm UV}}=-{\frac{\beta_1}{\beta_2}}$. Therefore we write $%
\beta(\lambda)=\beta_2\lambda^2 (\lambda-\lambda_{{\rm UV}})+{\cal O}%
(\lambda^4)$ and replace $\beta_1$ with $-\beta_2\lambda_{{\rm UV}}$
everywhere else. Finally, we compare the coefficient multiplying $\lambda_{%
{\rm UV}}^4$ in the expression of $a_{{\rm UV}}-a_{{\rm IR}}$ with
Hathrell's result. The numerical factor ${\frac{1}{12}}$, instead of ${\frac{%
1}{2}}$, is due to the different powers of $\lambda$ appearing in the beta
function and is crucial for the test.

\subsection*{Fourth-loop-order checks}

We can check agreement to the fourth-loop order using the calculations done
by Jack and Osborn in ref. \cite{jack}. Again, we have to merge these
results with some basic formulas of Hathrell's \cite{hathrell} to extract
the precise expression for $a^{\prime}$, which is not given explicitly in 
\cite{jack}. We do not give here the complete derivation, but provide a
vucabulary that allows the reader to surf on the various references and
notations. Unfortunately, the various pieces of the puzzle are distributed
in many different papers. For concreteness, we treat the case of a gauge
field theory.

Our $a^{\prime}$ is 
\[
-720 (4\pi)^2 (c^{\prime}+\bar c(\alpha)-\sigma(\alpha))= -720 (4\pi)^2
(c-\sigma) 
\]
in Hathrell's notation (this $c$ has nothing to do with our $c$). $%
c^{\prime} $ denotes the arbitrary additive constant. Hathrell proves that $%
c(\alpha)$ and $\sigma(\alpha)$ are related by the formula $\sigma=-\alpha {%
\frac{\partial \bar c}{\partial\alpha}}$. The quantity $\beta_c=-\sigma\beta$
coincides with the $\beta_c$ of \cite{jack}. The notations for the coupling
constants are as follows (\cite{jack}$\rightarrow$\cite{hathrell}): $g^i={%
\frac{1}{g^2}}\rightarrow {\frac{1}{4\pi\alpha}}$, $\beta^i=-{\frac{2}{g^3}}%
\beta(g)\rightarrow -{\frac{1}{4\pi\alpha}}\beta(\alpha)$.

$\beta_c$ is written as ${\frac{1}{8}}\left( \chi^{{\rm a}%
}_{ij}\beta^i\beta^j-\beta^i {\frac{\partial X}{\partial g^i}}\right)$ in 
\cite{jack} and $\chi^{{\rm a}}_{ij}={\frac{g^6}{4}}\chi^{{\rm a}}(g)$. The
explicit expression of $\chi^{{\rm a}}(g)$ (related to our function $f$ -
see below) is given in the second line of formula (5.12) of ref. \cite{jack}.

We find therefore 
\[
\sigma={\frac{\alpha}{8}}{\frac{\partial X}{\partial \alpha}}-{\frac{\pi}{8}}
\alpha\beta\chi^{{\rm a}}. 
\]
Denoting the total flow $k_{{\rm UV}}-k_{{\rm IR}}$ of a generic quantity $k$
with $\Delta k$, we have $\Delta \sigma={\frac{1}{8}}\Delta\left(\alpha {%
\frac{\partial X}{\partial \alpha}}\right)$. On the other hand, 
\[
\Delta\bar c=\int _{{\rm IR}}^{{\rm UV}}{\rm d}\alpha\, {\frac{\partial \bar
c}{\partial\alpha}}=-\int _{{\rm IR}}^{{\rm UV}}{\rm d}\alpha\, {\frac{\sigma%
}{\alpha}}=-{\frac{1}{8}}\Delta X-{\frac{\pi}{8}} \int^{{\rm IR}}_{{\rm UV}}%
{\rm d}\alpha\,\beta\chi^{{\rm a}}. 
\]
Therefore we can write 
\begin{equation}
\Delta a^{\prime}=90(4\pi)^2 \Delta \left( X+\alpha{\frac{\partial X}{%
\partial \alpha}}\right)+90\pi (4\pi)^2 \int^{{\rm IR}}_{{\rm UV}}{\rm d}%
\alpha\,\beta\chi^{{\rm a}}.  \label{blib}
\end{equation}
Now, we learn from formula (29) of \cite{osb} that 
\[
X+\alpha{\frac{\partial X}{\partial \alpha}}=-2\chi^{{\rm a}%
}_{ij}\beta^ig^j+\beta^i {\frac{\partial X^{\prime}}{\partial g^i}}, 
\]
for a certain regular function $X^{\prime}$ (called $Y$ in \cite{jack}).
This suffices to assure that the first term on the right-hand side of (\ref
{blib}) vanishes. Therefore we recover our formula for $\Delta a^{\prime}$
once we identify $f$ with $-90 \pi\chi^{{\rm a}} (4\pi)^2\alpha$. Using
(5.12) of \cite{jack} we see that this identification agrees with our
previous third-loop-order results ($f=45 \dim G+{\cal O}(\alpha)$).

On the other hand, we have 
\[
\Delta a=360(4\pi)^2 \int_{{\rm IR}}^{{\rm UV}}{\rm d}\alpha\, {\frac{%
\partial \tilde\beta_{{\rm b}}}{\partial \alpha}}= -{45\pi}(4\pi)^2\int_{%
{\rm UV}} ^{{\rm IR}}{\frac{{\rm d}\alpha}{\alpha}} \beta \chi^{{\rm g}%
}\alpha. 
\]
Now $\chi^{{\rm g}}=-2\chi^{{\rm a}}$ up to the fourth-loop order (see
formula (5.12) of \cite{jack}). This is the analogue of Hathrell's key
relation (formula (5.27) of \cite{hathrell}), used for the three-loop
checks. We conclude that the identification $\Delta a=\Delta a^{\prime}$ is
consistent to the fourth-loop order included, as we wished to show.

According to the references that we have used, the extension of the three-
and four-loop agreement to all orders is not trivial. In saying that the
higher order effects will conspire to satisfy our statement we are making a
strong claim. Pursuing this check to even higher orders would be desirable
and is not out of reach, given that exact formulas exist in supersymmetric
theories. The fifth-loop-order correction to $a$ ``just'' needs the
four-loop beta function \cite{four} and the three-loop expression of $f$.

\subsection{The Casimir effect}

The identification $a^{\prime}=a$ allows us to give an unambiguous
expression for the Casimir effect on a given manifold ${\cal M}$. The
derivation, that we do not repeat here, is performed by mapping the manifold 
${\cal M}$ into a conformally equivalent manifold ${\cal M}^{\prime}$ where
the effect is known. We refer to the papers of Bl\"ote, Cardy and
Nightingale \cite{blote} and Affleck \cite{affleck} for details. For
example, on a cylinder of radius $r$ the formula for the vacuum energy $E_0$
reads in our notation, using the results of Cappelli and Coste \cite{coste}, 
\[
E_0={\frac{1}{360}}{\frac{a}{r}},~~~~~~~~~~~~~~~~~~~~~~~~ \left( E_0=-{\frac{%
1}{12}} {\frac{c}{r}}\,\,\,\,\, {\rm in\, two \, dimensions.}\right) 
\]
For $a^{\prime}$ generic the expression reads 
\[
E_0={\frac{1}{1440}}{\frac{3a+a^{\prime}}{r}}, 
\]
and the shift of $a^{\prime}$ can be seen as a shift in the vacuum energy $%
E_0$. Quantum irreversibility can be reformulated as a statement on the
vacuum energy, $E_{{\rm UV}}\geq E_{{\rm IR}}\geq 0$.

\subsection{``Proper'' beta function and coupling constant}

Our formula, as we have stressed already, gives a natural geometrical
interpretation of quantum irreversibility, which turns out to be
quantitatively measured by the invariant area of the graph of the beta
function between the critical points. At intermediate scales $\mu$, the
quantity $a[\alpha(\mu)]$ knows about the area spanned by the part of graph
up to the scale $\mu$ (see Fig. 1). There is a universal cell of unit area
for each massless degree of freeedom. The number of massless degrees of
freedom disappearing between two given energy scales is equal to the area of
the graph of the beta function included between those scales.

Using the results of the previous sections, we can introduce a notion of
covariance related to scheme dependence and define a ``proper'' coupling
constant $\bar{\alpha}$ and beta function $\bar \beta(\bar \alpha)$. We
observe that the function $f(\alpha)$ is a metric in the space of couplings.
Precisely, when there are more couplings and $\Theta=\beta_i{\cal O}_i$ we
have 
\begin{equation}
\langle \Theta (x)\ \Theta (y)\rangle ={\frac{1}{15\pi ^{4}}}\,\,\frac{%
\beta_i[\alpha (t)]f_{ij}[\alpha (t)]\beta_j[\alpha (t)]}{|x-y|^{8}}%
~,~~~~~~~~~~{\rm for}~~x\neq y.  \label{tetatetaij}
\end{equation}
and the matrix $f_{ij}$ is positive definite. The total $a$-flow is then
expressed in the form 
\[
a_{{\rm UV}}-a_{{\rm IR}}=-\int_{\ln\alpha _{{\rm UV}}}^{\ln\alpha _{{\rm IR}%
}}{\rm d}\ln\alpha _{i}\ \beta _{j}(\alpha )f^{ij}(\alpha ); 
\]
$f^{ij}$ is a sort of ``target'' metric for the map $t\rightarrow \alpha
_{i}(t)$. This map is the path connecting $\alpha _{{\rm UV}}$ with $\alpha
_{{\rm IR}}$ in the space of couplings, and it is in general
scheme-dependent, as well as the metric $f^{ij}(\alpha )$ and $\beta _{i}$.
The integral is reparametrization-invariant. In this context
reparametrization invariance is precisely scheme independence.

\vskip .5truecm 
\let\picnaturalsize=N

\ifx\nopictures Y\else{\ifx\epsfloaded Y\else\fi
\global\let\epsfloaded=Y \centerline{\ifx\picnaturalsize N\epsfxsize
4.0in\fi \epsfbox{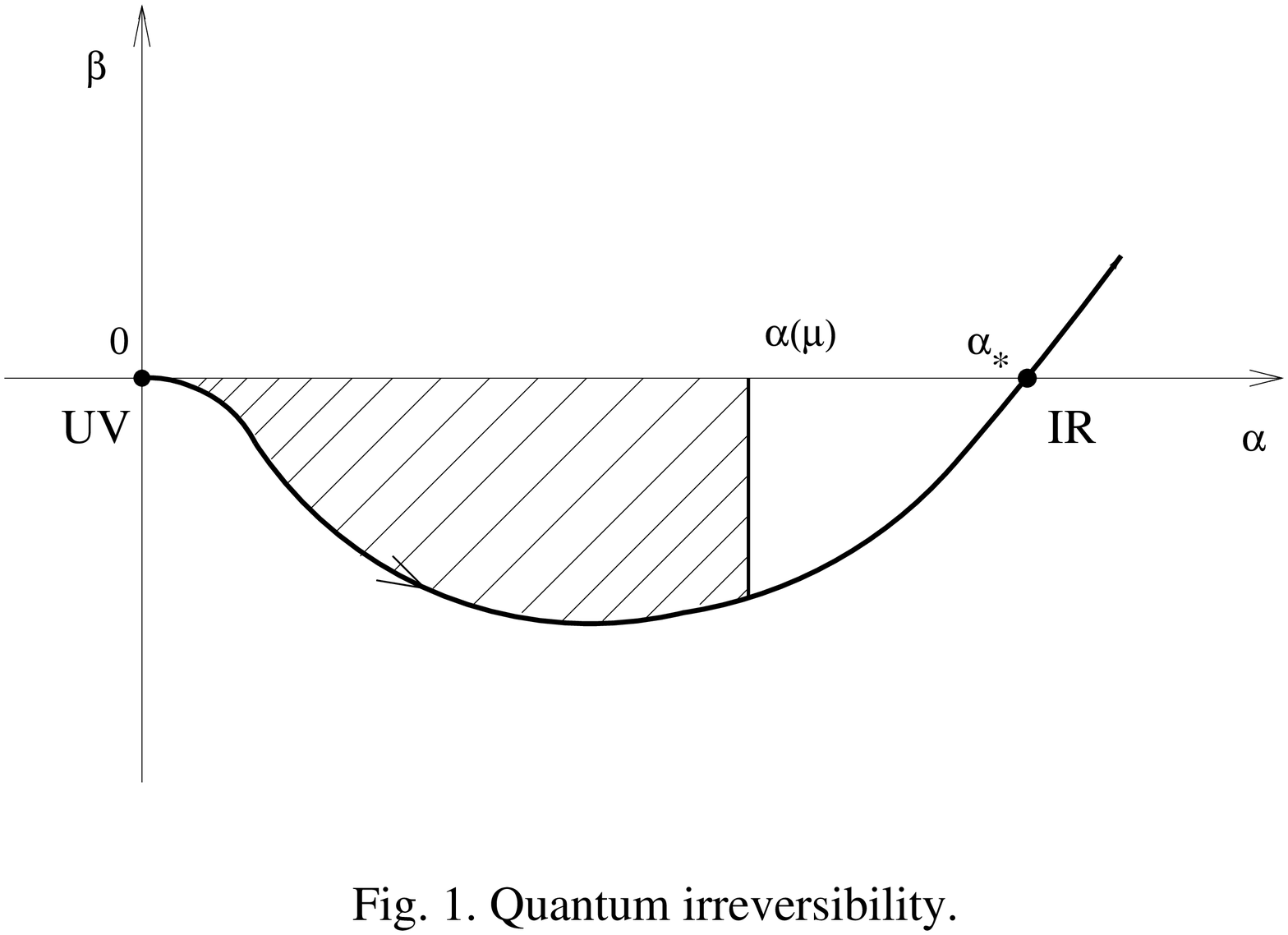}}}\fi

\vskip .5truecm

By definition, the proper coupling constant is the coupling constant for
which the metric is identically equal to the free-field (UV) value, $\bar{f}%
_{ij}(\bar{\alpha})=(f_{{\rm UV}})_{ij}$. Both $\beta_i $ and $f_{ij}$
depend on the scheme and in the ``proper'' scheme one can measure these
quantities in a universal way. 

Let us focus on the case of a single coupling constant $\alpha $, for
simplicity. By definition, we can write 
\[
\beta ^{2}(\alpha )f(\alpha )=\bar{\beta}^{2}(\bar{\alpha})f_{{\rm UV}}. 
\]
Moreover, we have 
\[
\bar{\beta}(\bar{\alpha})=\frac{{\rm d\ln }\bar{\alpha}}{{\rm d\ln }\mu }=%
\frac{{\rm d\ln }\bar{\alpha}}{{\rm d\ln }\alpha }\beta (\alpha ), 
\]
so that the formula relating the proper coupling constant to the starting
one, reads 
\[
\bar{\alpha}(\alpha )=\bar{\alpha}(\alpha _{0})\exp \left( \int_{\alpha
_{0}}^{\alpha }\frac{{\rm d}\alpha^{\prime}}{\alpha^{\prime}}\sqrt{{\frac{%
f(\alpha^{\prime})}{f_{{\rm UV}}}}}\right) . 
\]
$\bar \alpha$ is a power expansion in $\alpha$. 
An arbitrary integration constant survives in $\bar \alpha$ as a remnant of
scheme dependence. $\bar \beta$, instead, is uniquely fixed and universal.

For example, one can fix the integration constant at the IR fixed point
(which we assume to be an interacting conformal field theory), by setting $%
\alpha_0=\alpha_{{\rm IR}}$. 
There is no universal way to choose the overall factor $\bar{\alpha}(\alpha
_{{\rm IR}})$. If the IR fixed point is strongly coupled, then it is
reasonable to set $\bar \alpha_{{\rm IR}}=1$, by definition. Independently
of this value, one has $\bar{\alpha}_{{\rm UV}}=0$.


The loss of massless degrees of freedom along the renormalization group flow
is measured by the proper area of the graph of the beta function, i.e. the
area of the graph of the proper beta function, 
\[
a_{{\rm UV}}-a_{{\rm IR}}=-f_{{\rm UV}}\int_{\ln \bar{\alpha}_{{\rm UV}%
}}^{\ln \bar{\alpha}_{{\rm IR}}}{\rm d\ln }\bar{\alpha}\ \bar{\beta}(\bar{%
\alpha}). 
\]
To the lowest order in perturbation theory we have $\bar \alpha={\rm const.}%
\,\alpha+{\cal O}(\alpha^2)$. If we define, instead, $\bar \alpha$ as the
coupling for which the metric is exactly unity thgoughout the RG flow, so
that $\beta ^{2}(\alpha )f(\alpha )=\bar{\beta}^{2}(\bar{\alpha})$, we have $%
f(\alpha)/f_{{\rm UV}}\rightarrow f(\alpha)$ in the formulas above and the
transformation is no longer analytical, since around the free fixed point we
have $\bar{\alpha}(\alpha )={\rm const.}\,\alpha^{\sqrt{f_{{\rm UV}}}}$. For
example, in QCD $\ln \bar\alpha=\sqrt{45(N_c^2-1)}\ln \alpha+{\rm const.}$

We finally remark that the definition of proper coupling constant is valid
in any dimensions, even or odd, and in particular in three dimensions. In
odd dimensions, the integral (\ref{resu2}) is still a well-defined and
interesting physical quantity (it could be considered, by extension, the
effect of quantum irreversibility in odd dimensions), but there is no clear
definition of $a$ at criticality.




\section{Conclusions}

Several apparently unrelated facts suggest that the $a^{\prime }$-ambiguity
can be consistently removed by identifying $a^{\prime }$ with $a$. We have
analysed various arguments related with unitarity, renormalization and
positivity of the induced actions. In particular the
statement of positive definiteness 
of the Riegert action throughout the RG flow
is equivalent to the $a$-theorem.

The emerging picture clarifies several long-standing issues at the same
time, among which we recall the ambiguity of the term $\Box R,$ the
positivity of $a$, the decreasing behaviour of $a$ along the renormalization
group flow, the meaning of the Riegert action, the Casimir effect, the
conceptual differences between two and four dimensional quantum field theory.

%

The equality of the total flows of $a$ and $a^{\prime}$ gives an explicit
formula quantifying the effect of quantum irreversibility as the invariant
(i.e. scheme-independent) area of the graph of the beta function between the
fixed points. This formula can be checked explicitly to the fourth-loop
order in perturbation theory in all renormalizable models. There is a unity
``proper'' area associated with each massless degree of freedom disappearing
along the renormalization group flow.

This establishes the intrinsic relationship between renormalization (the
beta function), unitarity (absence of negatively normed states) and
irreversibility (disappearance of massless degrees of freedom along the
renormalization group flow), which we can schematically state as the
implication

\begin{center}
{\bf unitarity $+$ renormalization $\Rightarrow$ irreversibility.}
\end{center}

The interplay between unitarity and renormalization is better appreciated by
observing that in a running theory the requirement that there be no
negatively normed states naturally decomposes in two separate conditions:
the requirement that there be no negatively normed state at some reference
energy plus the requirement that no negatively normed state be generated
along the renormalization group flow. This interplay turns a simple set of
inequalities into the mentioned non-perturbative formula for the $a$-flow.

A by-product of our formula is an alternative, direct proof of the property 
\cite{noi} that $a$ is invariant with respect to marginal deformations: no
massless degrees of freedom can disappear along a trajectory with $%
\beta\equiv 0$. Given that also $c$ is invariant with respect to marginal
deformations \cite{noi}, a formula for the non-perturbative flow of $c$,
resembling (\ref{resu2}), should exist.

A further, non-trivial, implication is that along a non-trivial RG\ flow the
quantity $a$ strictly decreases. Therefore two conformal field theories with
the same $a $-values cannot be the critical points of an RG\ flow. This fact
was conjectured in \cite{high}. 

We have also traced the basic lines of an approach to the removal of
divergences in quantum field theory, according to which regularization and
renormalization are viewed as the violation and restoration of unitarity,
respectively. Negatively normed states are introduced to regularize and then
consistently removed to renormalize. In stressing the role of local terms
and running parameters in this context, as well as the issues related with
positive-definiteness of the induced effective actions, in particular
induced effective actions for external sources, this approach seems to be
more powerful than the usual one \cite{osterwalder} and deserves further
study {\sl per se}.

\vskip 1truecm

{\bf Acknowledgements.} I would like to thank A. Davydychev, P. Menotti, O.
Tarasov and F. Strocchi for correspondence, C. Angelantonj, I. Antoniadis,
J. F\"uchs, L. Girardello, G. Grunberg, G.~Mussardo, M. Porrati, R. Iengo
and A. Zaffaroni for discussions, and the Centre de Physique Th\'eorique de
l'Ecole Polytechnique, Palaiseau, for hospitality during the first stage of
this work.

\end{document}